\documentclass[preprint,showpacs,preprintnumbers,amsmath,amssymb]{revtex4}

\usepackage{graphicx}
\usepackage{dcolumn}
\usepackage{bm}

\newcommand{\be}{\begin{equation}}
\newcommand{\en}{\end{equation}}
\newcommand{\bea}{\begin{eqnarray}}
\newcommand{\ena}{\end{eqnarray}}

\begin{document}

\title{Curvaton reheating in  non-minimal derivative coupling to gravity: NO models}

\author{Ram\'on Herrera}
\email{ramon.herrera@pucv.cl} \affiliation{ Instituto de
F\'{\i}sica, Pontificia Universidad Cat\'{o}lica de
Valpara\'{\i}so, Av. Brasil 2950, Valpara\'{\i}so, Chile.}

\author{Joel Saavedra}
\email{joel.saavedra@pucv.cl} \affiliation{ Instituto de
F\'{\i}sica, Pontificia Universidad Cat\'{o}lica de
Valpara\'{\i}so, Av. Brasil 2950, Valpara\'{\i}so, Chile.}

\author{Cuauhtemoc Campuzano}
\email{ccvargas@uv.mx}
\affiliation{Departamento de F\'\i sica, Facultad de F\'\i sica
e Inteligencia Artificial, Universidad Veracruzana, 91000, Xalapa Veracruz,
M\'exico}

\date{\today}

\begin{abstract}
The curvaton reheating mechanism in a  non-minimal derivative
coupling to gravity for any non-oscillating (NO) model is studied.
In this framework, we analyze the energy density during the
kinetic epoch and we find  that this energy has a complicated
dependencies of the scale factor. Considering this mechanism, we
study the decay of the curvaton in two different scenarios and
also we determine the reheating temperatures. As an example the NO
model, we consider an exponential potential and we obtain the
reheating temperature indirectly from the inflation through of the
number of e-folds.

\end{abstract}

\maketitle

\section{Introduction}

Einstein's gravity has been considered for many years as the physics of the universe and the standard
model of the modern cosmology is based on this  theory. However, also it is well known if we are
searching for one complete description of the universe we need to use complements, in particular
we are using different kind of matter in order to do an  approximate description of the actual
universe. For example at the early universe we need to introduce the inflationary paradigm
in order to solve some problem  of the standard model, and it is introduced the inflaton field in order
to have the adequate expansion, and perturbation as seed of the large structure of the
universe \cite{R1,R2}. From the current observation we need to introduce the dark energy in order to do
an approximation to the observational data and touch the property theoretical description of
our currently observations. We can continue doing this addition of elements or complement
to the standard cosmological model in order to solve problems in the prediction, in some cases,
or to fit with the more and more precise astronomical data \cite{R4,Ade:2015lrj}. Alternative, we can change the
background
of the theory and we could consider modification to the Einstein's theory of gravity, as was done
in tensor-scalar theories or Jordan-Brans-Dicke theory \cite{JBD}. On the other hand, we can use actions from
higher dimensional theories or string theory, as the effective theory in the low energy limit \cite{RS}.
Also, we could do the modification of the theory or the action, in a straightforward way,
using a variational principle more general, for example
{\small{
\begin{equation}
S=\int d{}^4x\sqrt{-g}\left\{F(R,R_{\mu \nu}R^{\mu \nu},R_{\mu \nu \lambda \rho}
R^{\mu \nu \lambda \rho},..)+ K(\phi,\partial_{\mu}\phi \partial^{\mu}\phi,
\Box^2 \phi,R^{\mu \nu}\partial_{\mu}\phi\partial_{\nu}\phi,..) -V(\phi)\right\},\label{action1}
\end{equation}}}
where $F$ and $K$ are arbitrary functions of the corresponding variables. This action implies different
consequences, and we need to have in mind the basic principles of physics and therefore gravity.
The consequences can be see direct in the equation of motions, we must ask a priori (according to
the basic principles), a covariant formulation of equation of motion, the dynamics it is driven by
second order differential equation, and it must satisfy the correspondence principle. However,
the non-linear function $F$ and $K$  provided the general invariant that can meet at least two of
this requirement, unfortunately we must deal with higher order differential equation as equation of
motion. Of course here we have matter described by a scalar field, the way as this was introduced
in Einstein's gravity through the minimal coupling, between geometry and matter. In the action
(\ref{action1}),
we are considering in $K$ function the more general no-minimal coupling  between the scalar
field and gravity. Of course this new coupling modify the usual Klein-Gordon equation,  and
therefore the field equation for the scalar field is not longer a second order differential equation
in this general case, as an example of this higher dynamics see Refs.
\cite{Pulgar:2014cba, Amendola:1993uh, Capozziello:1999xt}. It is clear that the modification
of gravity in this way can be done by modification of the geometry
$F(R,R_{\mu \nu}R^{\mu \nu},R_{\mu \nu \lambda \rho}R^{\mu \nu \lambda \rho},\dots$)
or modification of matter sector
$K(\phi,\partial_{\mu}\phi \partial^{\mu}\phi, \Box^2 \phi,
R^{\mu \nu}\partial_{\mu}\phi\partial_{\nu}\phi,\ldots)$, this last choice we would like to discuss
in more details. Currently there are a growing interest in the called Horndeski Lagrangian
\cite{Horndeski:1974wa} the more general scalar field Lagrangian with non-minimal couplings
between the scalar field and the curvature, and   at the same time producing second order motion
equations. In Ref. \cite{Sushkov:2009hk},
 was found that the equation of motion for the scalar field can be reduced to second order
differential equation, when it is kinetically
coupled to the Einstein tensor, $G^{\mu \nu}\partial_{\mu}\phi\partial_{\nu}\phi$, and in
Ref. \cite{Saridakis:2010mf} the author investigated the cosmological scenarios for this kind of
coupling. In this case the action is described by\cite{Sushkov:2009hk}
\begin{equation}
S=  \int d^{4}x \sqrt{-g} \left( \frac{R}{16\pi G}-\frac{1}{2}\left( g_{\mu\nu}-\frac{1}{M^2}
G_{\mu \nu}\right)\partial^{\mu}\phi\partial^{\nu}\phi-V(\phi)\right)+S_{matter},\label{action2}
\end{equation}
where  $g$  corresponds to the determinant of the space-time metric $g_{\mu\nu}$, $R$ is the Ricci scalar
and
$G^{\mu \nu} = R^{\mu \nu} -
\frac{1}{2}Rg^{\mu \nu}$ is the Einstein tensor. Here the parameter $M$ is a constant with
dimension  of mass,
  and $V(\phi)$
corresponds to the effective potential associated to the scalar
field $\phi$.  The  parameter $M^{-2}$ and its sign plays a
critical role in this type of theory. Recently in
Ref.\cite{Starobinsky:2016kua} was studied the screening Horndeski
cosmologies in which  the ghost-free cosmological solutions occur
if the parameter $M^{-2}<0$. However,  analyzing
 the  dynamical stability in these solutions was found that they
 are
 stable in the future but unstable in the past
(initial spacetime singularity). In Ref.\cite{gh} was suggested
that the parameter $M^{-2}>0$ in order to evade the ghost presents
in the model. Also in Ref.\cite{gh2} was shown that independently
of the value of $M^{-2}$  the model  does not present
instabilities. In this form, we mention  that the sign of $M^{-2}$
positive or negative is  still an  open issue in the literature,
 in particular for  the early universe. In the following we
will consider the value $M^{-2}>0$ in order to study the early
universe. However, we mention that  for negative values of the
parameter $M^{-2}$, we  would have to  add the condition $1\gtrsim
H^2/M^2$, in order to evade
possible imaginary quantities in our model.

 The equation of motions for the geometry from the variation of the metric $\delta g_{\mu \nu}$
\begin{equation}
G_{\mu\nu}=8\pi G \left(T_{\mu\nu}^{\phi}+T_{\mu\nu}^{matter}+\frac{1}{M^2}
T_{\mu\nu}^{derivatives}\right),\label{EoMg}
\end{equation}
where
$T_{\mu\nu}^{matter}$ is the usual energy momentum tensor for the matter,
$T_{\mu\nu}^{\phi}=\nabla_{\mu}\phi\nabla_{\nu}\phi-\frac{1}{2} g_{\mu \nu}(\nabla \phi)^2$
and the new component is given by
\begin{eqnarray}
T_{\mu\nu}^{derivatives}&=&-\frac{1}{2}\nabla_{\mu}\phi\nabla_{\nu}R+
2\nabla_{\alpha}\phi\nabla_{(\mu}\phi R^{\alpha}_{\nu)}+
\nabla^{\alpha}\phi\nabla^{\beta}\phi R_{\mu\alpha\nu\beta}+
\nabla_{\mu}\nabla^{\alpha}\phi \nabla_{\nu}\nabla_{\alpha} \nonumber \\
&&-\nabla_{\mu}\nabla_{\nu}\phi \Box \phi-\frac{1}{2}(\nabla \phi)^2G_{\mu \nu}+
g_{\mu \nu}\left(-\frac{1}{2}\nabla^{\alpha}\nabla^{\beta}\nabla_{\alpha}\nabla_{\beta}+
\frac{1}{2}(\Box \phi)^2-\nabla_{\alpha}\phi\nabla_{\beta}\phi R^{\alpha \beta}\right).
\nonumber
\end{eqnarray}
The variation of the action respect to the scalar field $\delta \phi$ gives its equation of motion
\begin{equation}
\left(g^{\mu \nu}+\frac{1}{M^2}G^{\mu \nu}\right)\nabla_{\mu}\nabla_{\nu}\phi=
\frac{dV}{d\phi}.\label{eomphi}
\end{equation}

In relation to the cosmological consequences in non-minimally derivative
coupling to gravity were studied in Refs.\cite{Daniel:2007kk,Mas}.  Also, the
theory of the
 density perturbation  in the early universe with this non-minimally derivative
coupling was analyzed in Ref.\cite{Pe}.

On the other hand,
the reheating  of the universe is a procedure
in which the scalar field or  inflaton field  is converted into the standard model particles\cite{R15}.
During reheating of the universe, the best part of
the matter and radiation of the universe are created  via the decay of the scalar field
or inflaton.  Specifically, an important quantity known as the reheating
temperature, $T_{reh}$ can be found during this process. This quantity
is associated  to
 the temperature of the universe when the radiation epoch begin. A lower
 bound by the reheating temperature  from the Big Bang
 Nucleosynthesis (BBN) $T_{reh_{BBN}}\gtrsim 10^{-22} m_p$, where $m_p$ is the Planck
 mass, was obtained in Ref.\cite{BBN}.  Also, an upper bound by the reheating
 temperature arrives from the energy scale at the end of the inflationary period
 and
 is given by  $T_{reh}\lesssim 10^{-3}m_p$.  We mention that the
 first Bayesian constraints on the single field inflationary reheating epoch
from Cosmic Microwave Background  data
 was obtained in Ref.\cite{Re}, see also Ref.\cite{Re2}.

In this respect, afterward  of the inflationary period, the inflaton field
 experiments coherent oscillations at the bottom of an effective potential. In
 this form,
 an essential part in   the mechanism of reheating  are
the oscillations of the inflaton field. Nevertheless, there is in
the literature some models where the effective potential  does
not have a minimum and then the inflaton field does not oscillate. Therefore,
 the mechanism of reheating does not work.  In the literature, these models or
 those potentials that does not have a minimum
  are known  as non-oscillating (NO) models\cite{R16,R17}.

 For this type of NO model, the first mechanism of
reheating  was
 the gravitational particle production, nevertheless this mechanism becomes
inefficient, see  Refs.\cite{R21,R22}. The instants
preheating is  another mechanism for the NO model. The   instants
preheating  incorporates an interaction
between two scalar fields; the inflaton and another field \cite{R16}.
Another  alternative proposal to the reheating of the universe  in this type of  NO model, is the introduction
of curvaton field $\sigma$ \cite{R23}.  It is well known also that the curvaton field explains  the observed large-scale
 adiabatic density perturbation during the early universe \cite{R23}. In this
 respect,  the adiabatic density
perturbation  is produced from the curvaton field and not from the
inflaton field. In this framework, the adiabatic density
perturbation is originated   afterward inflation epoch and  from
an initial condition associates  to an isocurvature
perturbation\cite{R26}. Following  Ref.\cite{M1}, we assume   the
curvaton hypothesis, in which the observed value of the power
spectrum  the inflaton field ${\cal P}_{\zeta_\phi}$ is taken to
be less than the power spectrum  the curvaton ${\cal
P}_{\zeta_\sigma}$. Nevertheless, we mention that  in
Ref.\cite{M5} was considered  that
 the power spectrum  generated by both  fields  are important.
   Another
 important characteristics of the curvaton is that its energy density is
 subdominant while the inflation takes place and becomes dominant when the
 inflation finish. However, the curvaton survives to the expansion of the inflationary
 epoch.  In this respect, the curvaton reheating occurs when the curvaton decays  after or
 before dominate its energy density.  We mention that in Ref.\cite{Feng:2013pba} was studied
 a curvaton model, in which the curvaton has a nonminimal derivative coupling to
 gravity, see also Ref.\cite{Feng:2014tka}.

 In the framework of a  non-minimal derivative coupling to gravity
 we would
 like to introduce the curvaton field as a mechanism of reheating for any effective potentials that does  not
 minimum i.e., NO models. Therefore, the main aim of this paper
 is to carry out  the curvaton field into the non-minimal derivative coupling to
 gravity for NO models
 and see what consequences we may derive.
 The outline of the paper
 is as follow: in section II we give a brief review of the non-minimal derivative coupling inflationary epoch.
 In section III  we analyze  the kinetic epoch for non-minimal derivative coupling. In section IV we study the
 dynamic of the
 curvaton field.
 Section V describes the
 curvaton decay after its domination.  In section VI we analyze  the decay of the curvaton field
 before it dominates the expansion of the universe. In section VII we study a specific example of NO model, where we consider
 an exponential potential.
 At the end, in
 section VIII includes our conclusions.

\section{ non-minimal derivative coupling to gravity:  inflationary    epoch \label{secti}}
In this section we will briefly review of  the inflationary epoch in
the framework  the a  non-minimal derivative coupling to gravity.

\subsection{Inflationary epoch: A review }
In order to describe the  non-minimal derivative coupling inflationary
model, we start with the corresponding field equations that must
satisfy the scalar field in a flat Friedmann-Robertson-Walker
(FRW) background. From the action (\ref{action2}) we get
\begin{equation}
3H^2=\rho_\phi=\rho_\phi ^{kin}+\rho_\phi^V=
\left(1+\frac{9H^2}{M^2}\right)\frac{\dot{\phi}^2}{2}+V(\phi),
\end{equation}
and
\begin{equation}
\left(1+\frac{3H^2}{M^2}\right)\ddot{\phi}+3H\left(1+\frac{3H^2}{M^2}
+\frac{2\dot{H}}{M^2}\right)\dot{\phi}+V'(\phi)=0,
\label{e2}
\end{equation}
where $H:=\dot{a}/a$ is the Hubble parameter, $a=a(t)$ is the
scale factor,
 and $V' := \partial V/ \partial \phi$. Here the
kinetic energy density is defined as
 $\rho_\phi ^{kin}=(1+\frac{9H^2}{M^2})\frac{\dot{\phi}^2}{2}$ and the energy
 density associated to potential energy is given by  $\rho_\phi^V=V(\phi)$.
The dots denote derivative with respect to the cosmological time $t$, and we
shall use units such that $8\pi G=8\pi/m_p^2=1$.

Throughout inflation the energy density associated with the scalar
field is of the order of potential energy density, and dominates
over the kinetic energy, i.e., $\rho_\phi^V\gg\rho_\phi^{kin}$,
then the Friedmann equation can be written
as\cite{Tsujikawa:2012mk}
\begin{equation}
3H^2\simeq \rho_\phi ^V= V(\phi).\label{FI}
\end{equation}
 Here we note that during inflation the condition
$\rho_\phi^V\gg\rho_\phi^{kin}$ or equivalently $2V(\phi)\gg
\dot{\phi}^2(1+9H^2/M^2)$ coincides with first slow roll
approximation analyzed in Ref.\cite{Matsumoto:2015hua}. In this
form, the standard condition $V(\phi)\gg\dot{\phi}^2$ is modified
in the inflationary scenario  of non-minimal derivative coupling
to gravity.

On the other hand, the universe can  undergo a stage of
accelerated expansion only if $\ddot{a}>0$, and this condition is
model-independent, otherwise the gravity decelerates the
expansion.

In this form,  considering that during inflation $H^2>\dot{H}$ (
or equivalently $\ddot{a}>0) $ and neglecting  the acceleration of
the scalar field, the equation of motion associated to the scalar
field given by Eq.(\ref{e2}), reduces to\cite{Tsujikawa:2012mk}
\begin{equation}
3H\left(1+\frac{3H^2}{M^2}\right)\dot{\phi}+V'(\phi)\simeq 0,
\label{FII}
\end{equation}
and the velocity of the scalar field $\dot{\phi}$ becomes
\begin{equation}
\dot{\phi}=-\frac{V'}{\sqrt{3V}}\,\left(1+\frac{V}{M^2}\right)^{-1}.\label{g1}
\end{equation}
Here we have used Eq.(\ref{FI}). We mention  that the  high friction limit is
characterize by the condition $H^2\gg M^2$. Also, different  inflationary models
 in this limit was developed in Ref.\cite{Yang:2015pga} and numerical simulations in Ref.\cite{Tsujikawa:2012mk}.
 Also we mention that this condition of high friction limit
suggests the addition of   new conditions for the slow-roll
approximations given by $3\dot{\phi}^2/2M^2\ll 1$ and
$3V(\phi)/8\gg 1$, as was shown in Ref.\cite{Yang:2015pga}. Here,
the authors found that  these conditions give rise to solution of
the type Little Rip scenario. In the following we will not
consider this high friction limit and we will study the early
universe in the framework of Ref.\cite{Tsujikawa:2012mk}.

By introducing the slow-roll parameter $\epsilon$, we get
\begin{equation}
\epsilon=-\frac{\dot{H}}{H^2}\simeq\frac{V'^2}{2 V^2 (1+V/M^2)} .
\end{equation}
Now considering that inflation ends when the slow-roll parameter
$\epsilon=1$ (or equivalently $\ddot{a}=0$), then we can find
 the value of the potential $V(\phi=\phi_e)=V_e$ at the end of
inflation.

On the other hand, the number of e-folds $N_*$ is determined by
$N_* = \int_{t_*} ^{t_e} H(t')dt'$, and can be written as
\begin{equation}
N_* =-\int_{\phi_*}
^{\phi_e}\,\frac{V}{V'}\,\left[1+\frac{V}{M^2}\right]d\phi
.\label{nn}
\end{equation}
In the following, the subscripts $'*'$ and $'e'$ are used to indicate the time when
the cosmological scale leaves the horizon during inflation and the end of the
inflationary scenario, respectively.

\section{Kinetic epoch}
In this section we analyze  the kinetic epoch  of the inflaton  field.
 It is well known that when inflation has finished  the
model into the
 `kinetic epoch' (or `kination', for simplicity) \cite{Ag1}.  The kinetic epoch occurs at the end of inflation when
almost all the energy density of inflaton field is kinetic energy. However, we mention  that the
kinetic epoch does not take place immediately afterward of the inflationary
epoch \cite{Ag2}.

 By  assuming that during this epoch
 the kinetic energy
 $\rho_\phi ^{kin}
> \rho_\phi ^V \Leftrightarrow $
$(1+\frac{9H^2}{M^2})\dot{\phi}^2/2>V(\phi)$ and considering that
 the term
 $V'=\partial V/\partial
 \phi$,
 is very small  compared to the non standard friction term and the acceleration of the scalar
 field in the field Eq.(\ref{e2}), then   the field
equations during this epoch reduce to

\begin{eqnarray}
3H^2 = \rho_\phi ^{kin}\simeq \left(1+\frac{9H^2}{M^2}\right)\frac{\dot{\phi}^2}{2}
, \label{H2}
\end{eqnarray}
and
\begin{eqnarray}
\left(1+\frac{3H^2}{M^2}\right)\ddot{\phi}+3H\left(1+\frac{3H^2}{M^2}+
\frac{2\dot{H}}{M^2}\right)\dot{\phi}\simeq 0,\label{ddot}
\end{eqnarray}
respectively.

From Eq.(\ref{ddot}) we find a first integral  given by
\begin{equation}
\dot{\phi}=\frac{a_k ^3(1+3H_k
^2/M^2)}{a^3(1+3H^2/M^2)}\dot{\phi_k},\label{dphi}
\end{equation}
and corresponds to the velocity of scalar field during the kinetic
epoch. In the following, the subscription `k', labels the
different quantities at the starting of this epoch.

Combining  Eqs.(\ref{H2})  and (\ref{dphi}), we obtain that during
the kinetic epoch, the Hubble parameter in terms of the scale
factor  results
\begin{equation}
H^2(a)=H^2_{k}\left (\frac{F(a)}{F(a_k)}\right ),\label{ha}
\end{equation}
where the function $F(a)$ is given by
$$
F(a)=\frac{2^{1/3}\,(\frac{9A
\,a_k^6}{a^6}+1)}{3\left[B(a)+\sqrt{4(\frac{9A\,a_k^6}{a^6}+1)^3+B(a)^2}\,\right]^{1/3}}
+\frac{\left[B(a)+\sqrt{4(\frac{9A\,a_k^6}{a^6}+1)^3+B(a)^2}\,\right]^{1/3}}{32^{1/3}}-\frac{2}{3},
$$
in which
$$
B(a)=27\frac{A\,a_k^6}{a^6}+18\left(1-\frac{3A\,a_k^6}{a^6}\right)-16,
\,\,\;\;\mbox{with}\;\;\; A=\frac{\dot{\phi}^2_k
(1+3H^2_k/M^2)^2}{2M^2}.
$$

From the Friedmann equation given by  Eq.(\ref{H2}), the energy
density or kinetic energy $\rho_\phi^{kin}=\rho_\phi^{kin}(a)$,
can be written as
\begin{equation}
\rho_\phi^{kin}(a)=\rho_\phi ^k \left(\frac{a_k}{a}\right)^6
\left(\frac{1+3H_k ^2/M^2}{1+3H^2/M^2}\right)^2
\left(\frac{1+9H^2/M^2}{1+9H_k ^2/M^2}\right),\label{rho}
\end{equation}
where  the Hubble parameter $H=H(a)$ is given by Eq.(\ref{ha}) and
$H(a=a_k)=H_k$.

\begin{figure}[th]
{\includegraphics[width=4.in,angle=0,clip=true]{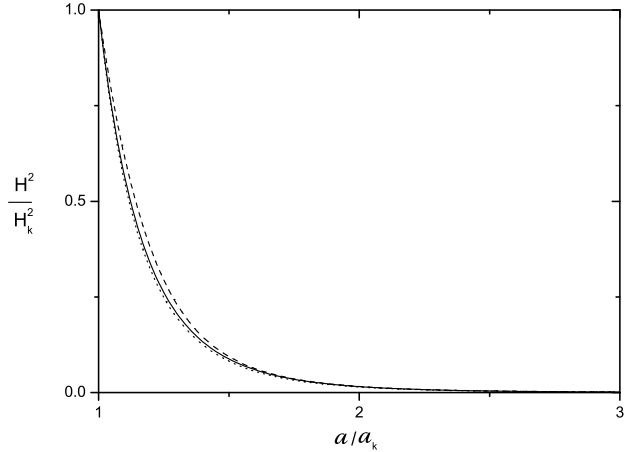}}
{\vspace{0.0 cm}\caption{ Evolution   of the square Hubble
parameter $H^2/H^2_k$ versus the scale factor $a/a_k$, for
different values of the dimensionless  parameter $A$ during the kinetic epoch.
The dot and dashed lines are for $A=0.1$ and $A=1$, respectively. The solid line
corresponds to the standard kinetic epoch in GR, in which $H\sim
a^{-3}$.
 \label{fig1}}}
\end{figure}

In Fig.(\ref{fig1})  we show the evolution of dimensionless square
Hubble parameter  $H^2/H^2_k$ versus the scale factor $a/a_k$, for
different values of the  dimensionless parameter $A$ during the
kinetic epoch. Here we consider the solution given by Eq.(\ref{ha}). In this plot  we analyze   two different values of the
parameter $A$ and also we consider the specific case of the standard kinetic epoch in
General Relativity (GR). In particular, the dot and dashed lines are for the specific
values of $A=0.1$ and $A=1$, and the solid line corresponds to the
standard kinetic epoch,  where $H\sim a^{-3}$. From this  plot we
note that when we decrease the parameter $A\rightarrow 0$, the
Hubble parameter during the kinetic epoch presents  a small
displacement with respect to the value of $H\sim a^3$. Also, we observe that the incorporation of the
new parameter $M$  gives us a freedom that allows us to change the
standard scenario of the kinetic epoch in GR.


\section{The dynamic of the curvaton  \label{sectii}}

In this section we study the dynamic of the curvaton field
 $\sigma$, through different epochs. From the dynamic of the
field $\sigma$, we  can find the constraints upon the parameter in
our model in order to obtain a viable model. For the dynamic,  we
assume  that the curvaton field satisfies the
Klein-Gordon equation with a scalar potential $U(\sigma)$ given by
$U(\sigma)=\frac{m^2\sigma^2}{2}$, where the parameter $m$ corresponds to the
curvaton mass.

Firstly, we assume that the inflaton field coexists with the
curvaton during the inflationary scenario. However, we consider
that the energy density associate to the curvaton field $\rho_\sigma$ is lower
than the energy density of inflaton field, i.e., $\rho_\sigma \ll
\rho_\phi$, such that the inflaton field $\phi$ always drives the
inflationary expansion of the early universe. In the next
scenario, the curvaton presents oscillations at the minimum of its
effective potential $U(\sigma)$. In this respect, the dynamic of
the  energy density of curvaton field evolves as a
non-relativistic matter and the expansion of the universe is even
dominated by the inflaton field. Finally, in the last scenario the
curvaton field decays into radiation, and then we recovered the
Big-Bang model.

During the inflationary expansion, is considered that the curvaton
mass  $m\ll H_e$, which means domination of the inflaton field over
curvaton, for more detail see Refs.\cite{dimo,postma,cdch}.
However, in the kinetic epoch the Hubble parameter decreases until
that its value becomes approximately to the curvaton mass, i.e.,
$H\simeq m$. From this condition and considering  Eq. (\ref{ha}),
we get
\begin{equation}
\frac{m^2}{H^2_{k}}\simeq\,\frac{F(a_m)}{F(a_k)},\label{mh}
\end{equation}
here the subscript $'m'$ stands the quantities evaluated at time
when the curvaton mass, $m\sim H$.

As it was commented above, we  considered  that the inflationary
epoch is only  driven by the inflaton field, and in order  to
prevent that the field curvaton produces an inflationary
expansion, we consider that the energy density of the inflaton
field at the times when $m\sim H$ becomes
$\rho_{\phi}|_{a_m}=\rho_{\phi}^{m}\gg\rho_{\sigma}$.  Over
inflation period, there is not substantial changes of the
effective potential, and then the energy density
$\rho_{\phi}^{m}\sim H^2\sim m^2\gg\rho_{\sigma}\sim U(\sigma_e)
\sim\,U(\sigma_*)$, resulting
\begin{equation}
\frac{m^2\sigma_*^2}{2\rho_\phi^{m}}=\frac{\sigma_*^2}{6}\ll\,\,1\,,\label{pot}
\end{equation}
or equivalently $\sigma_*^2\ll 6$. Here we note that the above
condition for the value  $\sigma_*$ coincides with the obtained in
Ref.\cite{R21}.

On the other hand, we note that  at the end of inflation the energy density of
the inflaton becomes subdominant over the energy of the curvaton
field, i.e. $V_e\gg U_e$. In this way, considering  Eq.
(\ref{pot})
 the ratio between the potential energies  can be written as

\begin{equation}
\frac{U_e}{V_e}=\frac{m^2\sigma_*^2}{6 H_e^2}\ll\,1\,\,\,\mbox{or
equivalently}\,\,\,\,\frac{m}{H_e}  \ll 1\label{u}.
\end{equation}
Here we note that the above inequality  gives a lower bound for
the curvaton mass $m$.

Since the Hubble parameter decreases during the expansion of the
Universe, then the mass of the curvaton field becomes significant
wherewith  $m\simeq H$,  and therefore its energy decays
$\rho_\sigma\propto a^{-3}$ i.e., as non-relativistic matter. In
this form,  we write
\begin{equation}
\rho_\sigma =
\frac{m^2\sigma_*^2}{2}\frac{a_m^3}{a^3}\label{c_cae}.
\end{equation}

In the following, we will consider the  decay of the curvaton
field in two different  scenarios; when  the curvaton field decays
after it dominates the expansion of the Universe and when the curvaton decays before it dominates.

\section{Curvaton Decay After Domination\label{sectiv}}
As we mentioned above  the curvaton field decays,  could take place
in two different possible scenarios. In the first scenario, the curvaton dominates
the cosmic expansion, i.e., the energy density of the curvaton
field $\rho_\sigma>\rho_\phi$. During the expansion there must be
an instant in which the energy densities of inflaton and curvaton
fields are equivalents, lets say, $a=a_{eq}$. Now from the
Eq.(\ref{rho})  and bearing in mind that
$\rho_\sigma\propto\,a^{-3}$, we have
\begin{eqnarray}
\left.\frac{\rho_\sigma}{\rho_\phi^{kin}}\right|_{a=a_{eq}}&=&\frac{m^2\sigma_*^2}{2}
\frac{a_m^3\;a_{eq}^3}{a_k^6\;\rho_\phi^k}
\left(\frac{1+3H_{eq}^2/M^2}{1+3H_k ^2/M^2}\right)^2
\left(\frac{1+9H_k ^2/M^2}{1+9H_{eq} ^2/M^2}\right)\nonumber\\
&=&\frac{m^2\sigma_*^2 a_m^3 a_{eq}^3}{6\;H_k^2 \; a_k^6}
\left(\frac{1+3H_{eq}^2/M^2}{1+3H_k ^2/M^2}\right)^2
\left(\frac{1+9H_k ^2/M^2}{1+9H_{eq} ^2/M^2}\right)=1
\label{equili},
\end{eqnarray}
where   we have used the relation $3\,H_k^2=\rho_\phi^k$, and also
the Hubble parameter $H(a=a_{eq})=H_{eq}$, is defined as
$H_{eq}=H_k\,[F(a_{eq})/F(a_k)]^{1/2}$, see Eq.(\ref{ha}).

On the other hand, as  the decay parameter $\Gamma_\sigma$ is limited from the
nucleosynthesis and the Hubble parameter during this epoch is
$H_{nucl}\sim 10^{-40}$ (in units of $m_p$), then a lower bound for the parameter $\Gamma_\sigma$
given by $H_{nucl}\sim 10^{-40}<\Gamma_{\sigma}$.  From the other side,  the
condition $\rho_\sigma > \rho_\phi$ (curvaton decays after
domination), we require $\Gamma_\sigma < H_{eq}$. In this way, the
constraint upon the decay parameter $\Gamma_\sigma$, can be
written as $
10^{-40}<\Gamma_{\sigma}<H_k\,[F(a_{eq})/F(a_k)]^{1/2} . $

 In the following  we will  study the scalar perturbations
related with the curvaton field $\sigma$. In order to describe the
curvature perturbation from the curvaton field, we mention two
possible stages. Firstly,  the quantum fluctuations during the
expansion of the universe are transformed  into classical
perturbations which have a flat spectrum. Secondly,  afterward
inflation the perturbations from the
curvaton field  are transformed  into curvature
perturbations and it does not need information about the nature of
inflation.

While the fluctuations are inside of the horizon, these have the
same differential equation that the inflaton fluctuations, wherewith
 the amplitude $\delta\sigma_*$ is given by  $\delta\sigma_*\simeq
H_*/2\pi$. Typically, the dynamics of the curvaton fluctuations outside of
the horizon, are like the unperturbed curvaton field, and these
fluctuations  remain constant during the expansion of the
universe.

In this context,  the power spectrum $P_\zeta\sim 10^{-9}$
\cite{Ade:2015lrj}, at the time when the decay of the curvaton takes
place and  can be written as    \cite{ref1u}
\begin{equation}
P_\zeta\simeq \frac{H_*^2}{9\pi^2\sigma_*^2}\simeq
\frac{V_{*}}{27\pi^2\sigma_*^2}\sim 10^{-9}, \label{pafter}
\end{equation}
where we have used Eq.(\ref{FI}).

From Eqs. (\ref{equili}) and (\ref{pafter}) we write a range for
the coefficient $\Gamma_\sigma$ given by $10^{-40}<\Gamma_\sigma<H_{eq}$ at the time in which curvaton
field decays after domination results

\begin{equation}
10^{-40}<\Gamma_\sigma
<\frac{M}{3^{1/2}}\left[\frac{3C_1}{2}-1+\sqrt{(3C_1/2-1)^2+(C_1-1)}\right]^{1/2},
\label{ww}
\end{equation}
where the constant $C_1\geq\frac{8}{9}$, and  is defined as
$$
C_1=\frac{(1+3H^2_k/M^2)^2}{(1+9H^2_k/M^2)}\,\left[\frac{\,a_k^6}{a_m^3a_{eq}^3}\right]\,\left[\frac{6H^2_k}{m^2\sigma_*^2}\right]\simeq
(1+3H^2_k/M^2)\,\left[\frac{\,a_k^6}{a_m^3a_{eq}^3}\right]\,\left[\frac{162\pi^2H^2_k\,P_\zeta}{m^2\,V_*}\right].
$$

In this form, in the first scenario we find an upper limit for the
reheating temperature  $T_{reh}\sim
\Gamma_\sigma^{1/2}$ and then from Eq.(\ref{ww}), we get
\begin{equation}
T_{reh}<\frac{M^{1/2}}{3^{1/4}}\left[\frac{3C_1}{2}-1+\sqrt{(3C_1/2-1)^2+(C_1-1)}\right]^{1/4}.\label{T1}
\end{equation}

 On the other hand, assuming that  the  BBN temperature $T_{BBN}$ is approximately equal to  $T_{BBN}\sim 10^{-22}$,
and considering that  the reheating temperature $T_{reh}$ occurs before the BBN,  then the reheating
temperature satisfies, $T_{reh}>T_{BBN}$. In this way,  considering that
$T_{reh}\sim \Gamma_\sigma^{1/2}>T_{BBN}$ and Eq.(\ref{ww}), we have

\begin{equation}
(162\pi^2 H^2_k\,P_\zeta)\,\left(1+\frac{3H^2_k}{M^2}\right)\,\left(\frac{a^6_k}{a^3_m
a^3_{eq}}\right)\,
\frac{(1+9T^4_{BBN}/M^2)}{(1+3T^4_{BBN}/M^2)^2}>m^2\,V_*
\,.\label{c}
\end{equation}

However,  we note that the curvaton decays occurs    before the
electroweak scale, since   the baryogenesis is situated  below the
electroweak scale, then  the quantity
$V_*^{1/4}\sim \sqrt{m_{ew}\,m_p}\sim 10^{10.5}$ GeV, in which
the electroweak scale $m_{ew}\sim 1$ TeV \cite{S1,S2}.
In this way, the square of the Hubble parameter satisfied
\begin{equation}
H_*^2\simeq\frac{V_*}{3}\sim 10^{-32},\label{elec}
\end{equation}
recalled  that $8\pi/m_p^2=1$. Now we note that  if the curvaton
decays before the electroweak scale, then from Eqs.(\ref{c}) and
(\ref{elec})  we obtain  an upper limit for the mass of the curvaton field given by
\begin{equation}
10^{26}\,\left(1+\frac{3H^2_k}{M^2}\right)\,\left(\frac{a^6_k}{a^3_m
a^3_{eq}}\right)\,
\frac{H^2_k}{(1+3T^4_{BBN}/M^2)}>m^2\,. \label{ww2}
\end{equation}
Here, we have used that $P_\zeta\sim 10^{-9}.$

\section{Curvaton Decay Before Domination\label{sectv}}

In this section we regard that  the curvaton $\sigma$ decays
before it dominates the expansion of the universe. In this context,
the mass of the curvaton  $m$, is non-negligible when is
contrasted  with the Hubble parameter $H$, and then we can consider
that the curvaton mass $m\sim H$. On the other hand, if the curvaton field
decays at a time when $\Gamma_\sigma =H(a_d)=H_d$, where ` d'
denotes the quantities at the time when the curvaton decays, then
from Eq.(\ref{ha}) we have
\begin{equation}
\Gamma_\sigma=\;H_d=\;H_k\;\sqrt{\frac{F(a_d)}{F(a_k)}}.
\label{Gamm}
\end{equation}

In this scenario, the curvaton field $\sigma$ should decay after that the mass
of the curvaton
$m\sim H$, satisfying  the condition $\Gamma_\sigma<m$. However, also
 the curvaton field $\sigma$ should decay before that it
 dominate the   expansion of the universe,
in which   $\Gamma_\sigma>H_{eq}$. In this form, considering  Eq.(\ref{equili}) we get
\begin{equation}
\frac{M}{3^{1/2}}\left[\frac{3C_1}{2}-1+\sqrt{(3C_1/2-1)^2+(C_1-1)}\right]^{1/2}<\Gamma_\sigma<\,m.
\label{gamm2}
\end{equation}
 Recalled that the
curvaton field decays at the time when $\rho_\sigma<\rho_\phi$.

Following Refs.\cite{ref1u,L1L2} the Bardeen  parameter $P_\zeta$, is given by

\begin{equation}
P_\zeta\simeq
\frac{r_d^2}{16\pi^2}\frac{H_*^2}{\sigma_*^2}\,,\,\,\,\,\;\;\;\mbox{where}\,\,\,\,\,\;\;\;
r_d=\left.\frac{\rho_\sigma}{\rho_\phi}\right|_{a=a_d},
\label{pbefore}
\end{equation}
in which  the parameter $r_d$ corresponds to the ratio
between the curvaton and the inflaton energy densities, measured
at the time in which the curvaton decay takes place.

Considering that the energy density of the curvaton decays as
non-relativistic matter i.e., $ \rho_\sigma \propto a^{-3}$, and rewritten the energy
density $\rho_\phi$ as
$$
\rho_\phi(a)=\rho_\phi^k\;\left(\frac{a_k}{a}\right)^6\;\frac{K(a)}{K(a_k)},
$$
where the new functions $K(a)$ is defined as
$$
K(a)=\left(\frac{1+3H_{k}^2/M^2}{1+3\frac{F(a)H_k
^2}{F(a_k)M^2}}\right)^2 \left(\frac{1+9\frac{F(a)H_k
^2}{F(a_k)M^2}}{1+9H_{k} ^2/M^2}\right),
$$
then  the ratio $r_d$,  results

\begin{equation}
r_d=\left.\frac{\rho_\sigma}{\rho_\phi}\right|_{a=a_{d}}=\frac{m^2\sigma_*^2}{6}
\frac{a_m^3\;a_{d}^3}{H_k^2\, a_k^6} \frac{K(a_k)}{K(a_d)}\label{2esce},
\end{equation}
or equivalently using Eq.(\ref{Gamm}) the ratio $r_d$ can be rewritten as
\begin{eqnarray}
r_d=\frac{m^2\sigma_*^2}{6}\frac{a_m^3\;a_{d}^3}{H_k^2\, a_k^6}
\left(\frac{1+3\Gamma_\sigma ^2/M^2}{1+3H_k ^2/M^2}\right)^2
\left(\frac{1+9H_k ^2/M^2}{1+9\Gamma_\sigma ^2/M^2}\right)\label{rd}.
\end{eqnarray}

From Eqs.(\ref{pbefore}) and (\ref{rd}), we find that the parameter
$\Gamma_\sigma$  can be written as
\begin{equation}
\Gamma_\sigma\approx\frac{M}{\sqrt{3}}\left[\frac{24\pi H_k
^2}{m^2 H_*\sigma_*} \sqrt{P_\xi} \left(\frac{a_k ^6}{a_m ^3 a_d
^3}\right) (1+3H^2_k/M^2)-1\right]^{1/2}.\label{s2}
\end{equation}

In this way, in the second scenario we find that the reheating
temperature using  Eq.(\ref{s2}) results
\begin{equation}
T_{reh}\sim \frac{M^{1/2}}{3^{1/4}}\left[\frac{24\pi H_k ^2}{m^2
H_*\sigma_*} \sqrt{P_\xi} \left(\frac{a_k ^6}{a_m ^3 a_d
^3}\right) (1+3H^2_k/M^2)-1\right]^{1/4}.\label{T2}
\end{equation}

Also, considering Eq.(\ref{gamm2}), we obtain that the condition for
the scalar field $\sigma_*$ becomes
\begin{equation}
\sigma_*<\frac{24\pi H_k ^2}{m^2 H_*\sigma_*} \sqrt{P_\xi}
\left(\frac{a_k ^6}{a_m ^3 a_d ^3}\right)
(1+3H^2_k/M^2)\,\left[\frac{3C_1}{2}+\sqrt{(3C_1/2-1)^2+(C_1-1)}\right]^{-1}\ll6.
\end{equation}
Here we have considered that $\sigma_*\ll6$, from the dynamic of the curvaton (see section IV).

\section{An example: Exponential potential\label{sectvi}}
In the following we study an exponential potential as an
example of NO model. The exponential potential is defined as

\begin{eqnarray}
   V(\phi)=V_0e^{-\alpha\phi}, \label{potent}
 \end{eqnarray}
where $V_{0 }$ and $\alpha$ are two positive parameters. This kind of potential was found
in power law inflation in which the scale factor $a(t)\propto
t^p$, where the exponent  $p>1$ \cite{matarrese}. Also the
exponential potential has been studied in the string theory and
tachyonic cosmologies \cite{250}. Another NO potentials can be found in
Ref.\cite{NO}.

From the exponential potential  and considering  Eq.(\ref{g1}), we
obtain that the scalar potential as function of the time (or $\phi(t)$), becomes
\begin{equation}
V(t)=V_0 e^{-\alpha \phi(t)}=\left[\sqrt{\frac{M^4}{4}
\left(C+\frac{\alpha^2}{2\sqrt{3}} t\right)^2+M^2}\;
 -\;\frac{M^2}{2}\left(C+\frac{\alpha^2}{2\sqrt{3}} t\right)\right]^2, \label{pot1}
\end{equation}
where the integration constant is defined as $C=\frac{e^{\alpha
\phi_0 /2} }{\sqrt{V_0 } } -\frac{1}{M^2}\sqrt{V_0}e^{-\alpha
\phi_0 /2}$, in which  $\phi(t=0)=\phi_0$.

From the slow-roll parameter $\epsilon$, we get
\begin{equation}
\epsilon=-\frac{\dot{H}}{H^2}=\frac{V'^2}{2 V^2 (1+V/M^2)}
=\frac{\alpha^2}{2(1+V/M^2)}.
\end{equation}
Now considering that inflation ends when the slow-roll parameter
$\epsilon=1$ (or equivalently $\ddot{a}=0$), then we find that the
value of the potential $V_e$ at the end of inflation results
\begin{equation}
V_e=M^2(\frac{\alpha^2}{2}-1),\label{vend}
\end{equation}
which implies that the parameter  $\alpha>\sqrt{2}$, since the
value of the potential $V_e
>0$. Also, we obtain that the number of e-folds $N_*$ results
\begin{equation}
N_* = \frac{1}{\alpha}\left[ \phi_e-\phi_*\right] +
\frac{1}{\alpha^2 M^2}
\left(V_*-V_e\right)=\frac{1}{\alpha^2}\left[\ln(V_*/V_e)+\frac{1}{M^2}(V_*-V_e)\right].\label{nn}
\end{equation}

\begin{figure}[th]
{\includegraphics[width=4.in,angle=0,clip=true]{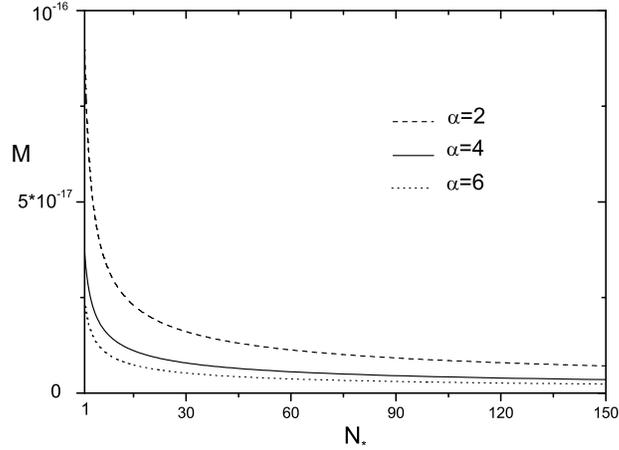}}
{\vspace{0.0 cm}\caption{ The  parameter $M$ as a function of the
number of e-folds $N_*$, for different values of the parameter
$\alpha$. The dot, solid and dashed lines are for the values
$\alpha=6$, $\alpha=4$ and $\alpha=2$. Here we have used that
$V_*=3\times 10^{-32}$.
 \label{fig2}}}
\end{figure}

In Fig.(\ref{fig2})  we show the  parameter  $M$ versus the number
of e-folds $N_*$, for different values of the parameter $\alpha$
associated to the exponential potential. Here we studied three
different values of the parameter $\alpha$. In order to write down
values that relate the parameter $M$ and the  number of e-folds,
we considering the relation given by Eq.(\ref{nn}). Also, we have
taken the value $V_*=3*10^{-32}$ from Eq.(\ref{elec}). In
particular, the dot, solid  and dashed lines are for the specific
values of $\alpha=6$ and $\alpha=4$, and $\alpha=2$, respectively.
We note that when we increase  the value of the parameter $\alpha$
(recall that $\alpha>\sqrt{2}$)  the number of
e-folds $N_*$ decreased and also the value of the parameter $M$.
Also from the plot we observe that the value of the parameter
$M<10^{-16}$ is well supported by the the number of e-folds $N_*\gtrsim 60$.

\begin{figure}[th]
{\includegraphics[width=4.in,angle=0,clip=true]{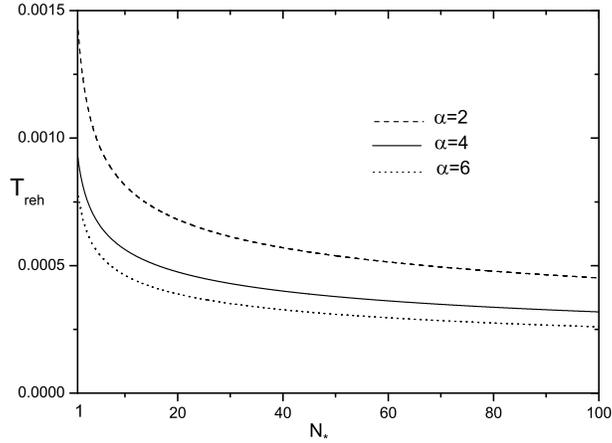}}
{\vspace{0.0 cm}\caption{ The reheating temperature $T_{reh}$ as a
function of the number of e-folds $N_*$, for different values of
the parameter $\alpha$ when the curvaton decays  after it
dominates the expansion of the universe. The dot, solid and dashed
lines are for the values $\alpha=6$, $\alpha=4$ and $\alpha=2$.
Here we have used the values $M=10^{-17}$, $m=10^{-20}$,
$H_k=10^{-17}$ and $a_k^2/(a_ma_{eq})=10^{-3}$.
 \label{fig3}}}
\end{figure}

\begin{figure}[th]
{\includegraphics[width=4.in,angle=0,clip=true]{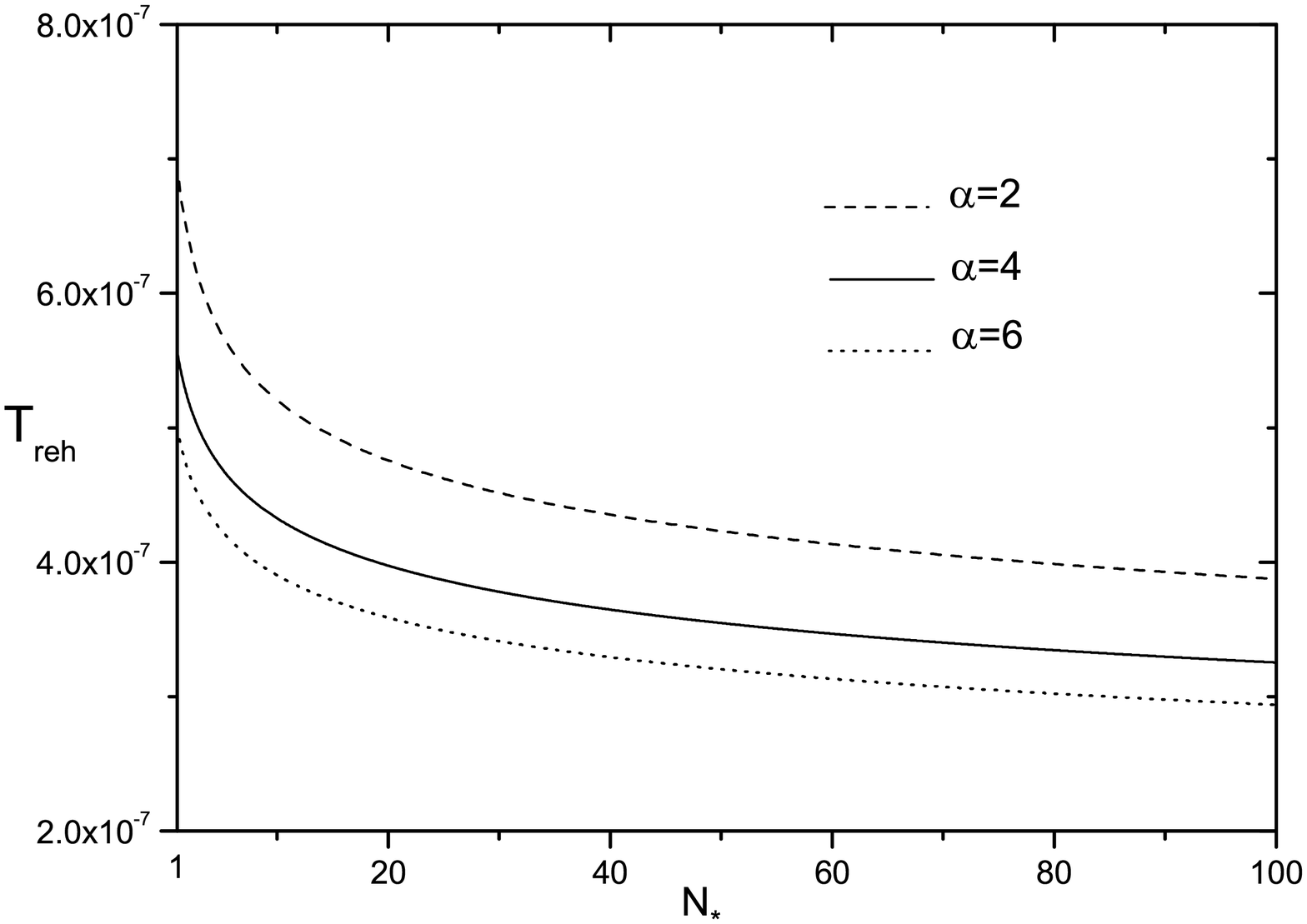}}
{\vspace{0.0 cm}\caption{ The reheating temperature $T_{reh}$ as a
function of the number of e-folds $N_*$, for different values of
the parameter $\alpha$ when the curvaton decays   before it
dominates the expansion of the universe. The dot, solid and dashed
lines are for the values $\alpha=6$, $\alpha=4$ and $\alpha=2$. As
before, we have used the values $M=10^{-17}$, $m=10^{-20}$,
$H_k=10^{-17}$, $\sigma_*=10^{-2}$, and
$[a_k^2/(a_ma_{d})]^3=10^{-10}$.
 \label{fig4}}}
\end{figure}

From  dynamics of the curvaton, we find that at the end of
inflation the energy density of the inflaton becomes subdominant
over the energy of the curvaton field, i.e. $V_e\gg U_e$. In this
way, considering Eqs.(\ref{pot}) and (\ref{vend}), the ratio
between the potential energies  can be written as

\begin{equation}
\frac{U_e}{V_e}=\frac{m^2\sigma_*^2}{6
H_e^2}\ll\,\frac{m^2}{H_e^2} = \frac{3 m^2}{M^2 (\alpha^2 /2-1)}
\ll 1\label{u},
\end{equation}
and then the ratio $m/M$, satisfied
\begin{equation}
\frac{m}{M}\ll\,\sqrt{\frac{(\alpha^2 /2-1)}{3}}.\label{mM}
\end{equation}
Here, we note that from the dynamic of the curvaton, we obtain  an
upper bound for the rate $m/M$.

In Fig.(\ref{fig3}) we show the reheating temperature $T_{reh}$ (in units of $m_p$) on
the number of e-folds $N_*$, when the curvaton field decays after
it dominates the expansion of the universe.  Here we have used
three different values of the parameter $\alpha$ associated to the
exponential potential, where the dot, solid and dashed lines are
for the values $\alpha=6$, $\alpha=4$ and $\alpha=2$. From
Eq.(\ref{T1}) we can obtain the reheating temperature $T_{reh}$ as
a function of the potential $V_*$,  i.e., $T_{reh}=T_{reh}(V_*)$
and together with Eqs.(\ref{vend}) and (\ref{nn}), we numerically
find the parametric plot of the curve $T_{reh}=T_{reh}(N)$.  This method to
determine the reheating  temperature in terms of the  number of e-folds $N_*$
during the evolution of the universe, was introduced in
Ref.\cite{Mielczarek:2010ag}.

 In
this plot we have considered the values $M=10^{-17}$, $m=10^{-20}$
from relation given by Eq.(\ref{mM}),
$H_*\simeq10^{-16}>H_k=10^{-17}$ see Eq.(\ref{elec}), and
considering that $a_k<a_m<a_{eq}$ then we have taken
$a_k^2/(a_ma_{eq})=10^{-3}$. We observe that the curves
$T_{reh}=T_{reh}(N)$ give an upper limit for the reheating
temperature, when the curvaton decays after domination in the case
of an exponential potential.  Also we note that when we increase
the value of $\alpha$, the reheating temperature $T_{reh}$
decreases to values $T_{reh}<10^{-3}$ for $N_*\simeq60$. Here we note that this
upper limit  for the reheating temperature is similar to the GUT scale, where
the temperatute $T_{reh_{GUT}}\lesssim 10^{-3}$ (in units of $m_p$). Also, we
observe that this upper limit in the $T_{reh}$,
is similar to that found in Ref.\cite{cdch}.

In Fig.(\ref{fig4}) we show the reheating temperature $T_{reh}$
versus  the number of e-folds $N_*$ when the curvaton field decays
before it dominates the expansion of the universe.  As before,  we
have used three different values of the parameter $\alpha$,
 where the dot, solid and
dashed lines are for the values $\alpha=6$, $\alpha=4$ and
$\alpha=2$. From Eq.(\ref{T2}) we can find the reheating
temperature $T_{reh}$ as a function of the potential $V_*$ and
together with Eqs.(\ref{vend}) and (\ref{nn}), we numerically
obtain the parametric plot  $T_{reh}=T_{reh}(N)$.  As before, in
this plot we have used the values $M=10^{-17}$, $m=10^{-20}$ and
$H_*\simeq10^{-16}>H_k=10^{-17}$. Also, we have considered that
$a_k^6/(a_ma_{d})^3=10^{-10}$ and  $\sigma_*=10^{-2}$. We note
from Fig.(\ref{fig4}) that when we increase the value of the
parameter $\alpha$, the reheating temperature $T_{reh}$ decreases
to values $T_{reh}<10^{-6}$ for $N_*\gtrsim60$. In particular for
the case in which $N_*=60$ and $\alpha=2$, we obtain that the
reheating temperature $T_{reh}\approx 4\times10^{-7}$, for the
value $\alpha=4$ corresponds to $T_{reh}\approx 3.5\times10^{-7}$,
and for the value $\alpha=6$ corresponds to $T_{reh}\approx
3\times10^{-7}$. It follows that one must increase the reheating
temperature by three orders of magnitude to have a $T_{reh}$ close
to the $T_{reh_{GUT}}$.

\section{Conclusions \label{conclu}}

We have analyzed  in  general form  and in detail the curvaton mechanism of reheating  into the NO
models
in the context of the non-minimal derivative coupling to gravity. In this framework,
we have considered  that
the curvaton field drives  the reheating the Universe as well as for the curvature perturbations.
Also, we have studied the kinetic epoch  in our model and we obtained  the
evolution of the Hubble parameter and kinetic energy expressed by Eqs. (\ref{ha}) and
(\ref{rho}), respectively.
In explaining  the curvaton reheating we have studied two possible scenarios: i) The
 curvaton field decays after it dominates the cosmic expansion of the universe and
 ii) the
 curvaton  decays before it dominates the expansion. During the first scenario,
 we have found an upper limit for the parameter $\Gamma_\sigma$ or
 equivalently an upper limit for the reheating temperature specified by
 Eq.(\ref{T1}). For the second scenario, we have obtained an approximate value
 for the temperature expressed by Eq.(\ref{T2}).

 As a specific example of  NO model, we have studied  an exponential potential.
 For this potential we have considered the method of constraining the reheating
 temperature indirectly from the inflationary period through the number of
 e-folds i.e., $T_{reh}=T_{reh}(N_*)$.  During the first scenario  when the
 curvaton  decays after it dominates the expansion,  we have found that for values of $\alpha>\sqrt{2}$, the
 reheating temperature is approximately $T_{reh}<10^{-3}$ (in units of $m_p$) as
 an upper bound. In the second scenario when the
 curvaton  decays before it dominates, we have obtained that the $T_{reh}<10^{-6}$ for values of $\alpha>\sqrt{2}$.
 We noted that these values for the temperatures are similar to those  found in
 Ref.\cite{cdch}.

\begin{acknowledgments}
 R. H.  and J. S.
were supported by the COMISION NACIONAL DE CIENCIAS Y TECNOLOGIA through FONDECYT Grant N$_0$
1130628. R. H. was partially supported by DI-PUCV Grant N$_0$ 123724.

\end{acknowledgments}


\end{document}